\documentclass[aps,prx,reprint,twocolumn,superscriptaddress,floatfix,nofootinbib]{revtex4-2}
\usepackage[english]{babel} 
\usepackage{adjustbox}
\usepackage{array}
\usepackage{amssymb}
\usepackage{amsfonts}
\usepackage{amsmath}
\usepackage{bm}
\usepackage{braket}
\usepackage{booktabs}
\usepackage{comment}
\usepackage{chngpage}
\usepackage{epsfig}
\usepackage{float}
\usepackage{graphicx}
\usepackage{mathtools}
\usepackage{mathrsfs}
\usepackage{mathdots}
\usepackage{multirow}
\usepackage[normalem]{ulem}
\usepackage{subfigure}
\usepackage{tabularx}
\usepackage{threeparttable}
\usepackage[toc]{appendix}
\usepackage{txfonts}
\usepackage{xcolor}
\usepackage{hyperref}
\usepackage[none]{hyphenat}
\hypersetup{
    unicode=false,     
    pdftoolbar=false,  
    pdfmenubar=true,   
    pdffitwindow=false, 
    pdfstartview={FitH},
    pdftitle={},    
    pdfauthor={Authors},     
    pdfsubject={},   
    pdfcreator={},   
    pdfproducer={}, 
    pdfnewwindow=true,
    colorlinks=true,
    linkcolor=black,
    citecolor=blue, 
    filecolor=magenta,
    urlcolor=blue
}

\newcommand{\thei}{t_{\mathrm{H}}}
\newcommand{\tauhei}{\tau_{\mathrm{H}}}
\newcommand{\tauplateau}{\tau_{\mathrm{p}}}

\newcommand{\tplateau}{t_{\mathrm{p}}}

\newcommand{\N}{N}

\newcommand{\hzero}{\hat{H}_{\mathrm{XY}}}

\newcommand{\Uh}{\hat{U}}
\newcommand{\Hh}{\hat{H}}

\graphicspath{{./figures/}}
\newcommand{\beginsupplement}{%
    \setcounter{table}{0}
    \renewcommand{\thetable}{S\arabic{table}}%
    \setcounter{figure}{0}
    \renewcommand{\thefigure}{S\arabic{figure}}%
    \setcounter{equation}{0}
    \renewcommand{\theequation}{S\arabic{equation}}%
    \setcounter{section}{0}
    \renewcommand{\thesection}{S\arabic{section}}%
    \setcounter{page}{1}
   }
   
\begin{document}

\title{Measuring Spectral Form Factor in Many-Body Chaotic \texorpdfstring{\\}{\space}
and Localized Phases of Quantum Processors}

\newcommand{\zju}{School of Physics, ZJU-Hangzhou Global Scientific and Technological Innovation Center, \\and Zhejiang Key Laboratory of Micro-nano Quantum Chips and Quantum Control, Zhejiang University, Hangzhou, China}

\author{Hang Dong} 
\thanks{These authors contributed equally.}
\affiliation{\zju}

\author{Pengfei Zhang}
\thanks{These authors contributed equally.}
\affiliation{\zju}

\author{Ceren B. Dağ} 
\affiliation{ITAMP, Harvard-Smithsonian Center for Astrophysics, Cambridge, MA 02138, USA}
\affiliation{ Department of Physics, Harvard University, Cambridge, Massachusetts 02138, USA}

\author{Yu Gao} 
\affiliation{\zju}
\author{Ning Wang} 
\affiliation{\zju}
\author{Jinfeng Deng}
\affiliation{\zju}
\author{Xu Zhang}
\affiliation{\zju}
\author{Jiachen Chen}
\affiliation{\zju}
\author{Shibo Xu}
\affiliation{\zju}
\author{Ke Wang}
\affiliation{\zju}
\author{Yaozu Wu}
\affiliation{\zju}
\author{Chuanyu Zhang}
\affiliation{\zju}
\author{Feitong Jin}
\affiliation{\zju}
\author{Xuhao Zhu}
\affiliation{\zju}
\author{Aosai Zhang} 
\affiliation{\zju}
\author{Yiren Zou}
\affiliation{\zju}
\author{Ziqi Tan}
\affiliation{\zju}
\author{Zhengyi Cui}
\affiliation{\zju}
\author{Zitian Zhu}
\affiliation{\zju}
\author{Fanhao Shen}
\affiliation{\zju}
\author{Tingting Li}
\affiliation{\zju}
\author{Jiarun Zhong}
\affiliation{\zju}
\author{Zehang Bao}
\affiliation{\zju}
\author{Hekang Li}
\affiliation{\zju}
\author{Zhen Wang}
\affiliation{\zju}
\author{Qiujiang Guo}
\affiliation{\zju}
\author{Chao Song}
\affiliation{\zju}
\author{Fangli Liu}
\affiliation{QuEra Computing Inc., 1284 Soldiers Field Road, Boston, MA, 02135, USA}
\author{Amos Chan}
\email{amos.chan@lancaster.ac.uk}
\address{Department of Physics, Lancaster University, Lancaster LA1 4YB, United Kingdom}
\author{Lei Ying}
\email{leiying@zju.edu.cn}
\affiliation{\zju}
\author{H. Wang}
\email{hhwang@zju.edu.cn}
\affiliation{\zju}

\date{\today}
\begin{abstract}
The spectral form factor (SFF) captures universal spectral fluctuations as signatures of quantum chaos, and has been instrumental in advancing multiple frontiers of physics including the studies of black holes and quantum many-body systems. 
However, the measurement of SFF in many-body systems is challenging due to the difficulty in resolving level spacings that become exponentially small with increasing system size.
Here we experimentally measure the SFF to probe the presence or absence of chaos in quantum many-body systems using a superconducting quantum processor with a randomized measurement protocol. 
For a Floquet chaotic system, we observe signatures of spectral rigidity of random matrix theory in SFF given by the ramp-plateau behavior. 
For a Hamiltonian system, we utilize SFF to distinguish the quantum many-body chaotic phase and the prethermal many-body localization. 
We observe the dip-ramp-plateau behavior of random matrix theory in the chaotic phase, and contrast the scaling of the plateau time in system size between the many-body chaotic and localized phases. 
Furthermore, we probe the eigenstate statistics by measuring a generalization of the SFF, known as the partial SFF, and observe distinct behaviors in the purities of the reduced density matrix in the two phases. This work unveils a new way of extracting the universal signatures of many-body quantum chaos in quantum devices by probing the correlations in eigenenergies and eigenstates.
\end{abstract}

\maketitle

\date{\today}

Spectral statistics is a powerful tool for analyzing quantum systems, as it captures the correlations between eigenenergies and reveals the universal features inherent to such systems.
It has served as a longstanding diagnostic of quantum chaos as described by the Bohigas-Giannoni-Schmidt conjecture~\cite{bohigas1984characterization}: A quantum system can be considered chaotic if its spectral statistics resemble those found in random matrix theory (RMT)~\cite{Mehta} at sufficiently small energy scales. Historically, spectral statistics and RMT have been applicable in a wide range of fields, including complex atomic nuclei~\cite{porter1956, brody1981nuclear}, number theory~\cite{montgomery1973number, keating2000random}, quantum chaos~\cite{bohigas1984characterization, berry1985semiclassical}, and quantum transport in mesoscopic systems \cite{Imry_1986, AltshulerShklovskii}.
Experimentally, spectral statistics have been probed by extracting the energy levels from the Fourier transform of a time-dependent correlation function~\cite{martinis2017spec} or from spectroscopy measurements~\cite{haq1982, frisch2014quantum, assmann2016quantum, zhou2010magnetic}, 
with emphases on nearest-neighbor level spacing.
However, such protocols are challenging for many-body systems due to the difficulty in resolving level spacings that become exponentially small with increasing system size.
Beyond spectral correlations, another important diagnostic of quantum chaos and thermalization, concerning eigenstate correlations, is the Eigenstate Thermalization Hypothesis (ETH)~\cite{deutsch1991quantum, Srednicki, Rigol2008}. The ETH postulates that in sufficiently narrow energy windows, the matrix elements of few-body operators in the energy eigenstate behave in a typical way as captured by the RMT, thereby providing an explanation of thermalization in isolated quantum systems through the understanding of correlations among eigenstates.
Spectral statistics and eigenstate correlations serve as two defining diagnoses for the presence (or absence) of quantum chaos and thermalization, and therefore their experimental measurements are of immense interest in the study of out-of-equilibrium dynamics in quantum many-body systems.

\begin{figure}
\centering
\includegraphics[width=\linewidth]{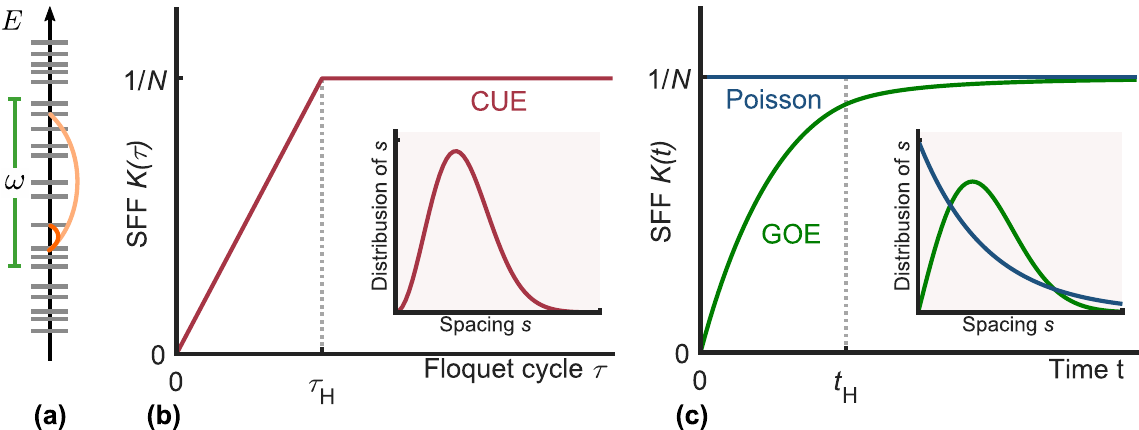}
\caption{
 {\bf Schematics of the SFF.} 
	(a) Spectral correlations concern the likelihood of finding two levels with a certain distance $\omega$, and it can be probed in the time domain by its Fourier transform known as the SFF. 
 (b) Floquet quantum chaotic systems without time reversal symmetry [c.f. Fig.~\ref{fig:floquet_fig}] display universal SFF behavior that can be captured 
 by random matrix ensembles known as the circular unitary ensemble (CUE). The CUE behavior displays a ramp followed by a plateau after the Heisenberg time $\thei$, and the behavior around $\thei$ is also often probed with reference to the Wigner-Dyson distribution in the inset, which is the nearest-neighbor energy spacing distribution found in the RMT. 
(c) Quantum chaotic systems with time reversal symmetry [c.f. Fig.~\ref{fig:Ham_Model}] display universal SFF behavior of the Gaussian orthogonal ensemble (GOE, in green), while the SFF of localized systems can be captured 
 by Poisson distribution (blue). The corresponding spacing distributions are given in the inset.  
 } \label{fig1}
\end{figure}

The \textit{spectral form factor} (SFF) is the Fourier transform of the two-level correlation function -- the probability of finding two eigenenergies with a certain distance in the energy spectrum [Fig.~\ref{fig1} (a)].
The SFF is arguably the simplest analytically tractable quantity to capture the \textit{long-range} universal spectral fluctuation of quantum systems, and consequently, it has been instrumental in multiple frontiers of physics, such as the semi-classical approach to quantum chaos~\cite{berry1985semiclassical, Sieber_2001, muller_2004, muller_2005},  black holes~\cite{kitaev2015simple, garciagarcia2016, Cotler_2017, Saad2019semiclassical}, many-body quantum chaos~\cite{chan2018solution, kos_sff_prx_2018, chan2018spectral, bertini2018exact}, and transition to prethermal many-body localization (MBL)~\cite{basko2006metal, oganesyan2007localization, chaoschallengeMBL, Prakash2021} and more \cite{Cotler_2017complexity, del_Campo_2017, ALTLAND201845, Liu_2018, liaogalitski2020, chan2020lyap, roy2020random, SUNTAJS2021168469, winer2022hydrodynamic, sfffilter2022delcampo, Winer_2022glass, santos2022}.
Formally, the SFF can be defined as
\begin{equation}\label{eq:sff_def}
K(t)
= \frac{1}{\N^2} \overline{\, \, 
\left| \mathrm{Tr}\, \Uh(t) 
\right|^2
}
= \frac{1}{\N^2} \overline{\sum_{a,b} e^{i(E_a-E_b)t}}
\;, 
\end{equation} 
where $\hat{U}$ is the time evolution operator of the system of interest, defined through either a time-independent Hamiltonian via $\Uh(t ):= \exp\left( -i\Hh t \right)$, or a time-periodic Hamiltonian via $\hat{U}(t= \tau T)= \hat{U}^\tau$ with Floquet operator $\Uh:=\mathcal{\hat{T}} \exp\left(-i\int_0^{T} \mathrm{d}t^{\prime} \, \hat{H}(t^{\prime})\right)$.
Here $\mathcal{\hat{T}}$ is the time-ordering operator. $T$  and $\tau$ are the Floquet period and number of Floquet cycles respectively. $\{ E_{a} \}$ are the eigenenergies or quasienergies of $\Hh$ and $\Uh$ respectively. $\N$ is the Hilbert space dimension, and $\overline{(\dots )}$ denotes the ensemble average over statistically-similar systems. We adopt the convention where $K(t)$ is normalized such that $K(0)=1$. 
For quantum many-body systems, a generalization of SFF called the \textit{partial spectral form factor} (pSFF)~\cite{ZollerSFF2021, Gong_2020,  Garratt2021prx} has been introduced by partitioning the many-body system, and was utilized to probe eigenstate correlations in addition to spectral statistics.
Crucially, both the SFF and the pSFF are defined in the time domain which provides an avenue to circumvent the experimental obstacles mentioned above.
Benefitting from the rapid development of controllable quantum simulators~\cite{Preskill2018nisq, quantumsim2021},
and protocols that utilize randomized sampling and repeated measurements~\cite{Elben_2022, ZollerSFF2021},
we measure the SFF and the pSFF in superconducting quantum simulators to diagnose signatures of quantum chaos and localization in periodically-driven and time-independent quantum many-body systems.

\begin{figure*}[!htbp]
\centering
\epsfig{figure=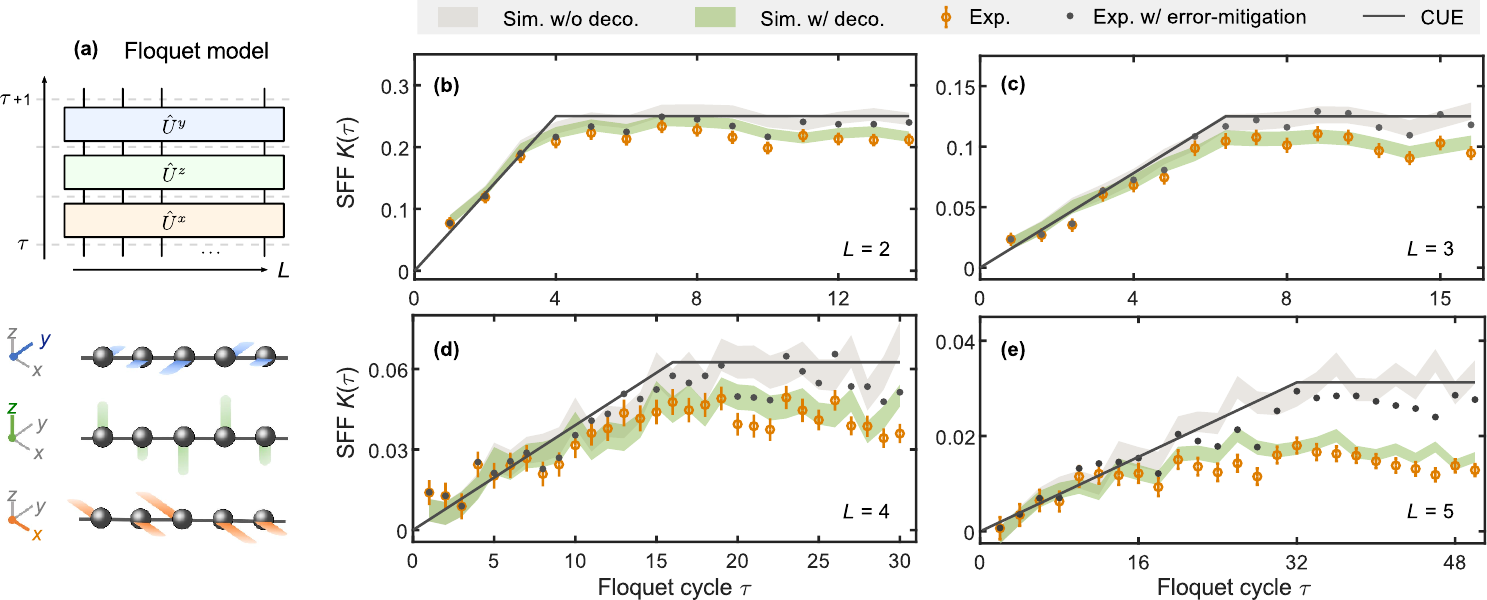,width=0.95\linewidth}
\caption{{\bf Experimental measurement of the SFF for a Floquet quantum many-body system, Eq.~\eqref{eq:model_flo}, in the chaotic phase.}
(a) Illustration of the Floquet system, with time-dependent driving along the $x,z,y$ axes within a single Floquet cycle.
(b-e) SFF of the Floquet system against the number of Floquet cycles with system sizes up to $L=5$ for 400 to 1000 realizations. 
The experimental data, error-mitigated data, simulations, and RMT prediction are plotted in circles, black dots, shaded areas, and lines respectively. 
}\label{fig:floquet_fig}
\end{figure*}

%

We utilize superconducting many-body quantum simulators which have one-dimensional arrays of individually controllable qubits with tunable nearest-neighbor couplings (see details in Supplementary Information (SI)). Specifically, we simulate the one-dimensional XY model, $ \hzero = J \sum_{m}^{L-1} \left( \hat{\sigma}_m^{+}  \hat{\sigma}_{m+1}^{-} + \hat{\sigma}_m^{-}  \hat{\sigma}_{m+1}^{+} \right)$, along with time-varying potential on each qubit of the form, $h_{m}^\alpha (t) \, \hat{\sigma}_m^\alpha $. Here, $J$ denotes the strength of tunable nearest-neighbor coupling. $L$ denotes the number of the qubits in the chain.
$\hat{\sigma}_m^{\alpha}$ denotes a Pauli matrix acting on the $m$-th qubit with $\alpha= x,y,z$, and $\hat{\sigma}^{\pm}_m := \left(\hat{\sigma}^{x}_m \pm i \hat{\sigma}^y_m\right)/2$. The driving $h_{m}^\alpha (t)$ of the $m$-th qubit along the $\alpha$-axis  can be individually and dynamically tuned, allowing us to simulate both many-body Floquet and time-independent Hamiltonian systems.

We first program one of the simulators to implement a Floquet (i.e. time-periodic) system, which is an ideal test bed for quantum many-body dynamics due to its simplicity. Such a model can be constructed so that all global symmetries and conserved quantities, including energy, are removed. Furthermore, unlike the case in time-independent systems, 
the behavior of SFF in Floquet models at early times is not affected by the inhomogeneities in the density of states. Our Floquet model with $L$ qubits is defined by [Fig.~\ref{fig:floquet_fig}a],
\begin{equation} \label{eq:model_flo}
\begin{aligned}
    \hat{H}(t) &= \left\{
    \begin{array}{llll}
    \hat{H}^{x}, \quad&\quad0  &\leq t/T < 1/3, \\
    \hat{H}^{z}, \quad&1/3&\leq t/T < 2/3, \\
    \hat{H}^{y}, \quad&2/3&\leq t/T < 1, 
    \end{array}
    \right.
\\
\hat{H}^{\alpha} &= \;  \hzero +\sum_{m=1}^{L} h_m^{\alpha} \hat{\sigma}_m^{\alpha},
\end{aligned}
\end{equation}
where the coupling strength $J$ in $\hzero$ is tuned to be $-5$~MHz. The local potentials $h_m^{\alpha}$ are randomly and independently sampled from a uniform distribution over a range of $[-W, W]$ with $W/2\pi = 5$~MHz. 
The periods $T$ are chosen to be $150~\text{ns}$ for $L=2,3$ and $90~\text{ns}$ for $L=4,5$  (see SI for details).
As alluded earlier, this Floquet model has the advantage of having no symmetries and conserved quantities, and the integrability is broken due to the driving.

For quantum chaotic systems with Hilbert space size $N$ and without symmetries, the behavior of SFF can be modeled by the circular (or Gaussian) unitary ensemble (CUE) in the RMT of $N$-by-$N$ unitary matrices. Correspondingly the SFF $K(\tau)$ at Floquet cycle $\tau$ displays a characteristic feature known as the ``ramp-plateau behavior''~ \cite{Mehta}, $K_{\mathrm{CUE}}(\tau)= {\tau}/{\tauhei N}$ 
for $0<\tau \leq \tauhei $ and $1/N$ for $\tau>\tauhei$, as depicted in Fig.~\ref{fig1} (b). Here  $\tauhei =  N $ is the analogue of the Heisenberg time in Floquet systems, which is a dimensionless quantity proportional to the inverse of the mean quasi-energy level spacing. The transition between the ramp and plateau at the Heisenberg time $\tauhei$ reflects the same physics captured by level repulsion, which is often probed with reference to the Wigner-Dyson nearest-neighbur energy level spacing distribution.

Fig.~\ref{fig:floquet_fig} (b-e) shows the experimental data on SFF of the Floquet model with system size $L$ changing from $2$ to $5$. 
After taking account of decoherence (see below), we see a good qualitative agreement between the experimental data and the numerical simulation. 
The ramp-plateau RMT behavior is apparent, especially for the smaller system sizes in early time when the decoherence effects are not prominent. 
We observe that the plateau time, $\tauplateau$, defined to be the the time when the plateau begins, approximately coincides with the Heisenberg time $\tauhei = N= 2^L$.
The observation of the ramp-plateau behavior is a compelling evidence that this system is in the chaotic phase.
Decoherence in the quantum processer lowers the value of $K(\tau)$, leading to deviations between the experimental measurement of $K(\tau)$ and the theoretical prediction $K(\tau)$ in the absence of decoherence. 

To understand the deviations, we simulate the experimental protocol including the decoherence parameters, where the qubit lifetimes $T_1$ and  spin-echo times $T_2$ have averages of $\sim 90 \,  \mu$s and $\sim 14 \, \mu $s respectively (see SI),
and find that the numerics is in qualitative agreement with the experimental data (Fig.~\ref{fig:floquet_fig}). 
Since $\tauhei$ scales 
exponentially with the system size $L$, the observation of the ramp-plateau behavior for larger system sizes are more prone to decoherence. 
Recovering the intrinsic SFF behavior of a quantum many-body system from the experimental data including decoherence is challenging, and currently is an open problem.
Nonetheless, we adopt two simple methods to mitigate the errors in the behavior of SFF in the presence of decoherence.
First, we apply a formula derived in \cite{ZollerSFF2021} under the assumption that the decoherence can be approximated by a global depolarization channel (see SI). 
Secondly, we rescale the experimental data with the ratio between the numerics of SFF in the presence and absence of decoherence (see Fig.~2). Even though these approximations on the decoherence are crude,
we find that both methods qualitatively recover the ramp-plateau behavior with plateau times that are consistent with our expectations.

Next, we program the 
simulator to implement a one-dimensional time-independent many-body Hamiltonian  defined by
\begin{equation} \label{eq:model_ham}
    \hat{H} = \hzero +h^x \sum_m^L \hat{\sigma}_{m}^x + \sum_{m}^L h^z_m \hat{\sigma}_m^z \;,
\end{equation}
where the coupling strength $J$ in $\hzero$ and tranverse field $h^x$ are configured at $-5$~MHz and  $2$~MHz respectively. The local potentials $h^z_m$ are randomly and independently sampled from a uniform distribution within the range $[-W, W]$. The value of $W$ determines the dynamical regime of the system --- whether it exhibits the characteristics of quantum chaotic or prethermal MBL systems. Although the stability of the MBL phase in the thermodynamic limit is currently under debate \cite{chaoschallengeMBL, abanin2021distinguishing}, our experimental system is a finite-size system, and MBL can exist as a metastable state. 
\begin{figure*}[htbp]
\centering
\epsfig{figure=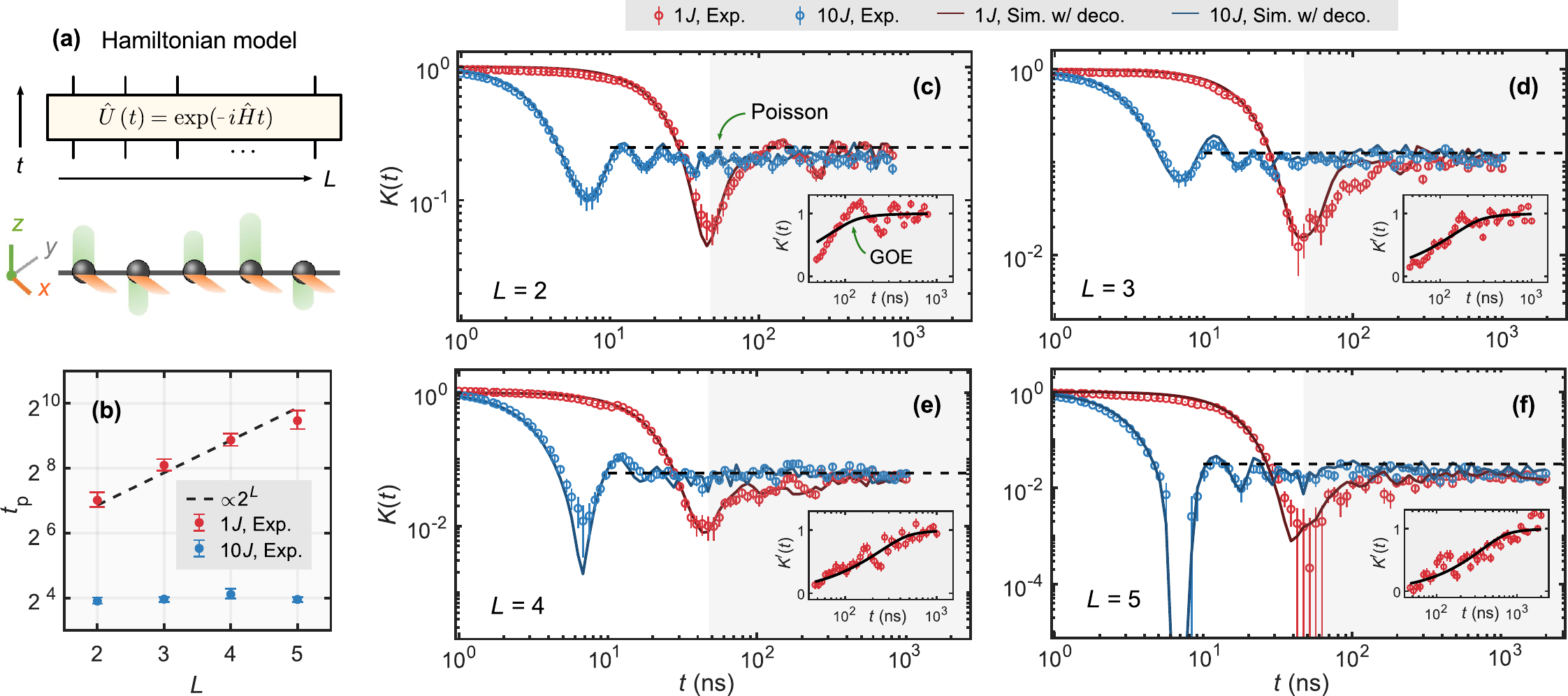,width=0.92\linewidth}
\caption{ {\bf Experimental measurement of the SFF for a Hamiltonian system, Eq.~\eqref{eq:model_ham}, in the chaotic and prethermal MBL phases.} 
(a) Illustration of the Hamiltonian system, which is simulated in the chaotic phase with $W=J$ and prethermal MBL phase with $W= 10J$ for 400 to 1000 realizations up to the system size $L= 5$. 
(b) The plateau time $\tplateau$ of the SFF increases exponentially in $L$ in the chaotic phase (red), but not in the prethermal MBL phase (blue). 
(c-f) SFF $K(t)$ of the Hamiltonian model in the chaotic phase (blue) and the prethermal MBL phase (red) for up to $L=5$.  The experimental data, simulations, and Poissonian statistics are plotted in circles, lines and dashed lines respectively. 
For each inset, we numerically fit the error-mitigated and normalized SFF data $K^{\prime}\left(t\right)$ with the GOE RMT behavior, for the domain in the shaded region of the main panel starting from $t= 48$~ns when the SFF has reached its lowest value.
See SI for details of the determination of $\tplateau$, the fitting and error mitigation procedures.}
\label{fig:Ham_Model}
\end{figure*}

In the presence of time-reversal symmetry with the time reversal symmetry operator squaring to identity, quantum chaotic systems display RMT spectral statistics as described by the Gaussian (or circular) orthogonal ensemble (GOE). The connected SFF\footnote{The \textit{connected SFF} is defined as $K_\mathrm{con}(t)= K(t) - 
\frac{1}{\N^2} \left| \overline{\sum_{a} e^{iE_a t}} \right|^2 $ where the initial non-universal dip behavior in $K(t)$ due to inhomogeneities in the density of states is removed.} of the GOE with $N$-by-$N$ random matrices is given by $K_{\mathrm{GOE}}(t)=\frac{1}{N}\left[ \frac{2t}{\thei} - \frac{t}{\thei} \ln \left(1+ \frac{2t}{\thei}\right) \right]$ before $\thei$, and $K_{\mathrm{GOE}}(t)= \frac{1}{N}\left[ 2 - \frac{t}{\thei} \ln \left(\frac{2 t/\thei+1}{2 t/ \thei-1}\right) \right]$ after $\thei$, which is proportional to the inverse of the mean level spacing [Fig. \ref{fig1} (c)]. 
Whereas for MBL and certain integrable systems, spectral statistics can be modeled by  Poisson distribution, and SFF quickly approaches $N^{-1}$~\cite{BerryTabor, Prakash2021} [Fig.~\ref{fig1} (b)].

The measurement results and numerical simulations of the SFF for time-independent Hamiltonian both in chaotic and prethermal MBL phases are shown in Figure~\ref{fig:Ham_Model} (c)-(f) for system sizes up to $L=5$. For both cases, for sufficiently large time, SFF will reach the plateau at the plateau time $\tplateau$.
For $W=J$, SFF displays an initial dip, a ramp, and a plateau with $\tplateau$ scaling exponentially in the system size. The experimental data are in qualitative agreement with the numerical simulation and theoretical prediction from the RMT, suggesting that the system is in the chaotic phase, and that $\tplateau$ can be identified as $\thei$ of the system.
For $W= 10J$, the SFF dips and then plateaus relatively quickly, at times much earlier than the case of $W=J$. Crucially, unlike the chaotic case, $\tplateau$ does not increase exponentially with system size [Fig.~\ref{fig:Ham_Model} (b)], which is consistent with the expectation of a prethermal MBL phase. 
For both cases, the theoretical behaviors from the RMT and the Poissonian distribution [Fig.~\ref{fig1}] are fitted after the initial dip in error-mitigated SFF data [see insets in Fig.~\ref{fig:Ham_Model} (c-f)], to avoid the early-time non-universal SFF behavior appearing due to the inhomogeneities in the density of states.
See SI for the details on the fits of the data and the determination of $\tplateau$. 

\begin{figure*}[!htbp]
\centering
\epsfig{figure=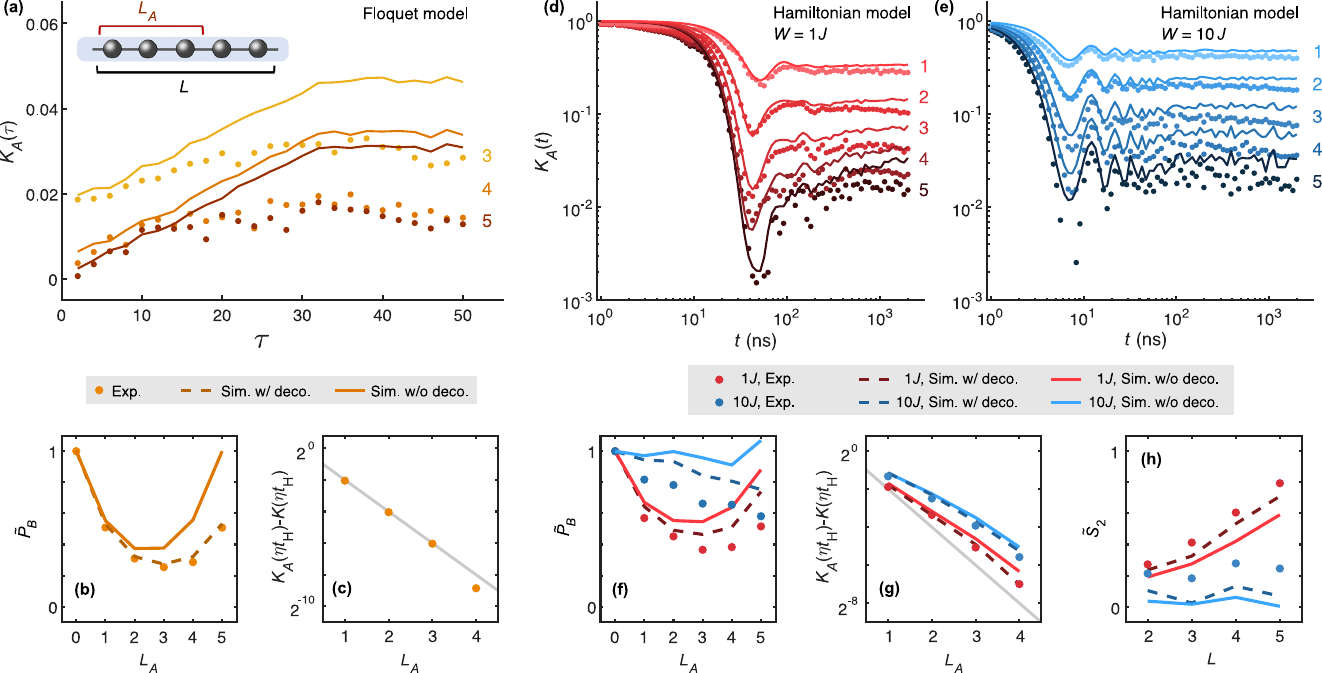,width=\linewidth}
\caption{\label{fig:PSFF}{\bf Experimental measurement of the pSFF for the Floquet system, Eq.~\eqref{eq:model_flo}, and the Hamiltonian system, Eq.~\eqref{eq:model_ham}.}
The experimental measurement of the pSFF for the Floquet system in the chaotic phase (a), the Hamiltonian system in the chaotic phase (d) and the prethermal MBL phase (e). The system size for all cases is $L=5$, and the integers towards the right of (a,d,e) denote the subsystem sizes $L_A \in [1,5]$.
We estimate the purity $P_B$ of the subsystem $B$ by measuring the long time behavior of pSFF, $\tilde{P}_B \equiv K_A(\eta \, \thei) N_A$ with $\eta=1.2$ for (b) Floquet and (f) Hamiltonian systems. 
We measure the dependence of the shift on $L_A$ in pSFF concerning the SFF by computing $K_A(\eta \, \thei) - K(\eta \, \thei)$
for the (c) Floquet and (g) Hamiltonian systems. The gray lines are given by $N_A^{-2}= 2^{-2L_A}$.
In (h), we plot the estimated 
half-system-size annealed average of second Renyi entropy $S_2$ as a function of $L$ by computing $\tilde{S}_2 \equiv -\log \tilde{P}_B$. 
}
\end{figure*}

The partial spectral form factor (pSFF) is a generalization of the SFF in quantum many-body systems that probes the correlations of eigenstates, in addition to the correlations of eigenenergies. Due to this feature, pSFF can detect the signatures of thermalization, (prethermal) localization, or other ergodicity-breaking mechanisms exhibited in eigenstate correlations. Specifically, for a subsystem $A$ and its complement $B$, the pSFF is defined as \cite{ZollerSFF2021, Gong_2020, Garratt2021prx}
\begin{equation}
K_A\left(t\right):= \frac{1}{N N_A} \overline{ \,
\mathrm{Tr}_B\left[\left| \mathrm{Tr}_A\Uh(t)
\right|^2\right]
} \; ,
\end{equation}
where $N_A$ 
denotes the Hilbert space dimension of subsystem $A$, and $\mathrm{Tr}_A$ is the partial trace of subsystem $A$. Note that $K_A\left( t \right)$ is normalized such that $K_A(0)=1$, similar to $K(t)$ above.
Experimentally, the pSFF can be accessed using the randomized measurement protocol identical to the one of SFF~\cite{ZollerSFF2021}, except that only the random measurements in the subsystem $A$ are taken, see Methods below. 
In Fig.~\ref{fig:PSFF} (a) and (d), we experimentally measure and numerically simulate the pSFF for the Floquet model and the Hamiltonian model in the chaotic phase, respectively.
After accounting the effect of decoherence, we observe the RMT ramp-plateau behavior in $K_A(t)$ with the qualitative agreement between experimental data and numerical simulations. 
We also observe a vertical shift in $K_A(t)$ that is dependent on subsystem size, see Fig.~\ref{fig:PSFF}  (c) and (g).
These qualitative features are consistent with the pSFF behavior for chaotic quantum many-body systems at sufficiently large time, such that the systems exhibit RMT behavior as given by~\cite{ZollerSFF2021},
$K^{\mathrm{RMT}}_{A}(t) = [N^2(N^2_A - 1)K(t) + (N^2-N_A^2)]/N_A^2(N^2 - 1) \approx K(t) + 1/N_A^2$, where the approximation is valid when $N_A, N/N_A \gg 1$.
In Fig.~\ref{fig:PSFF} (e), we measure the pSFF for the Hamiltonian model in the prethermal MBL phase. 
In contrast to the chaotic phase (as in the case of SFF), we observe that pSFF quickly reach the plateau with a plateau time $\tplateau$ that does not scale with the system size.
As in the chaotic case, pSFF for MBL displays a subsystem-size-dependent shift, but the shift in the MBL case is larger than the one in the chaotic phase due to differences in eigenstate correlations in the two cases, as reflected in the purity discussion below.
For all cases, by analyzing the data after the removal of the vertical shift, we observe that the plateau time $\tplateau$ of pSFF does not display dependency on the subsystem size $L_A$, and coincides with the $\tplateau$ of SFF, see analysis in the SI.

The purity of the reduced density matrix of a subsystem $B$, defined by $p_B:= \operatorname{Tr}_B \left[\rho_B^2 \right]$, quantifies the entanglement between the subsystem and its complement, and is related to the second Renyi entropy via $S_2(\rho_B):= -\log p_B$.
The averaged purity can be expressed in terms of the pSFF $K_A(t)$ for $t$ much larger than the Heisenberg time $\thei$ as 
$P_B \equiv \frac{1}{N} \overline{p_B} 
=\lim_{\thei/ t \rightarrow 0 } K_A(t ) \, N_A$. 
In practice, since experimental data is collected in finite time (with finite-depth circuits) and the decoherence is increasingly dominant at late times, we obtain an estimation of the purity, $\tilde{P}_B$, by measuring the pSFF (for both the chaotic and localized phases) at $\eta \, \thei$ with $\eta = 1.2$. 
In Fig.~\ref{fig:PSFF} (b) and (f), we present the experimental results with numerical simulations for the Floquet and Hamiltonian models respectively, which display decent agreement. 
We see a substantial difference in purity $\tilde{P}_B$ (consequently the estimation of the annealed average of second Renyi entropy) in the chaotic and localized cases, namely that $\rho_B$ is more mixed in the chaotic case compared to the localized case.
Note that in the presence of decoherence, $\tilde{P}_B$ obtained under the current protocol (dots in Fig.~\ref{fig:PSFF} (b) and (f), where only measurement data in subsystem $A$ are utilized) is not symmetric under the exchange of $L_A$ and $L-L_A$ \cite{Brydges_2019} unlike the numerical simulations without decoherence channels (solid lines).   
With the accessible system sizes, we observe in Fig.~\ref{fig:PSFF} (h) that the estimation of the half-system-size annealed average of second Renyi entropy, $\tilde{S}_2$, has a faster increase in the total system size in the chaotic case.
These results demonstrate that, as expected, subsystems share more quantum entanglement in chaotic quantum many-body systems than in prethermal MBL systems.
We have experimentally measured spectral form factors in quantum many-body superconducting processors, 
thereby demonstrating the effectiveness of such processors in probing the signatures of quantum chaos in spectral statistics and eigenstate correlations.
%
For the first time to our knowledge, we observe the long-range spectral rigidity in both time-independent and periodically-driven quantum many-body systems by measuring the random matrix theory predicted ramp-plateau behavior in spectral form factors. 
Further, utilizing both spectral form factor and partial spectral form factor as its generalization, we demonstrated the existence of a prethermal many-body localized regime in our time-independent setup and contrasted its eigenenergy and eigenstate correlations against those in the chaotic regime.
%
%
The experimental measurement of form factors opens up exciting directions in charting the dynamical signatures of many-body quantum systems in the laboratory, such as the universal behavior of the spectral form factor in earlier times than the onset of random matrix theory 
~\cite{chan2018spectral, Gharibyan_2018,Dag:2022vqb, yoshimura2023operator}, the emergence of random matrix theory universality in non-Hermitian systems~\cite{Shivam_2023}, the crossover 
between the chaotic and prethermal many-body localized regimes~\cite{chaoschallengeMBL}.  

\vskip 0.5cm

\noindent {\bf Acknowledgments:} The device was fabricated at the Micro-Nano Fabrication Center of Zhejiang University.  
We acknowledge the support of the National Natural Science Foundation of China (Grants No.~92365301, 92065204, 12174342, 12274368, 12274367, 12375021 and U20A2076), the National Key Research and Development Program of China (Grant No.~2023YFB4502600 and 2022YFA1404203), and the Zhejiang Provincial Natural Science Foundation of China (Grant No.~LR24A040002).
A.C. acknowledges the support from the Royal Society grant RGS{$\backslash$}R1{$\backslash$}231444, and the fellowships from the PCTS at Princeton University. 

\vskip 0.5cm

\noindent {\bf Author contributions:} F.L., L.Y., A.C. and C.D. proposed the ideas and laid out the theoretical background; Y.G., H.D., and P.Z. performed the numerical simulation in collaboration with A.C., C.D., and F.L.; H.D., P.Z., and Y.G. carried out the experiments and analyzed the experimental data under the supervision of H.W.; H.L. and J.C. fabricated the device supervised by H.W.; H.D., A.C., P.Z., L.Y., C.D., and H.W. co-wrote the manuscript; All authors contributed to the experimental setup, and/or the discussions of the results and the writing of the manuscript.


%

\clearpage

\beginsupplement
\onecolumngrid
\begin{center}
    \textbf{\large Supplementary material for ``Measuring Spectral Form Factor in Many-Body Chaotic and Localized Phases of Quantum Processors''}\\[.2cm]

\end{center}
\section{Experimental setup and protocol for the SFF measurement}

\begin{figure}[!htb]
\centering
\includegraphics[width=1.0\textwidth]{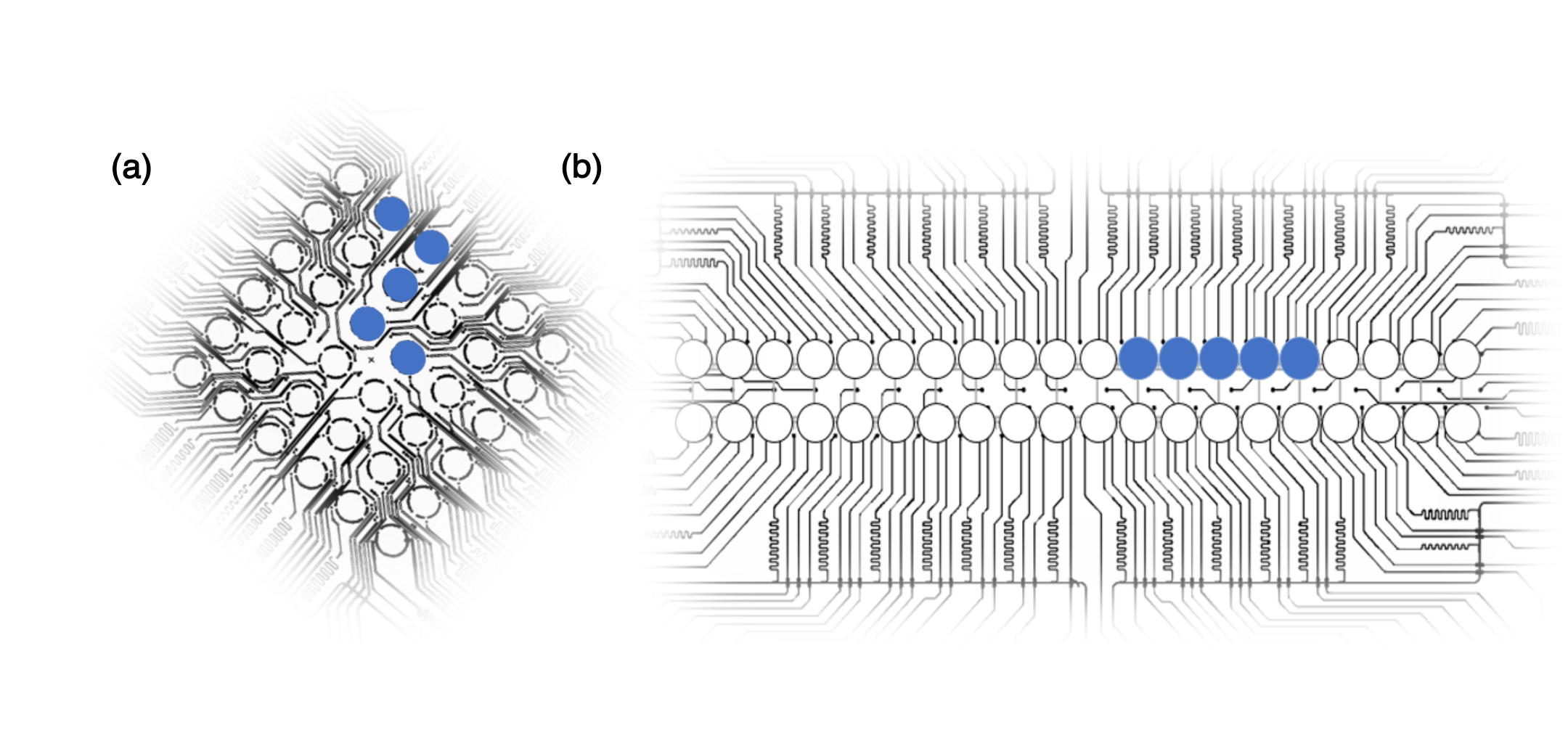}
\caption{\label{Fig1}{\bf Schematics of processor~\MakeUppercase{\romannumeral 1} (a) and processor~\MakeUppercase{\romannumeral 2} (b).} Processor~\MakeUppercase{\romannumeral 1} has a $6 \times 6$ qubit array, and processor~\MakeUppercase{\romannumeral 2} has a $2 \times 20$ qubit ladder. All qubits used in the experiment are depicted with blue filled circles.}
\label{fig:processors}
\end{figure}

Two flip-chip superconducting quantum processors, named processor~\MakeUppercase{\romannumeral 1} and processor~\MakeUppercase{\romannumeral 2} (Fig.~\ref{fig:processors}), are used to implement the Floquet and the Hamiltonian models, respectively.
These two processors have slightly different layouts, but we can select a chain of $L$ qubits featuring controllable nearest-neighbor couplings in both processors, with $L$ up to 5 for the experiments.
For both processors, each qubit has an individual flux (Z) control for statically and dynamically tuning its resonant frequency and a microwave (XY) control for arbitrary single-qubit XY rotations, based on which any single-qubit Clifford gates can be assembled~\cite{PhysRevA.90.030303}.
Two nearest-neighbor qubits directly connect to a coupler, which is also frequency-tunable, so that the effective qubit-qubit coupling can be switched from off to on by lowering the coupler's frequency from high to low.
The effective coupling strength can be dynamically modulated in a range of $\left[-15, -3\right]$~MHz for both quantum processors. 
Basic performance parameters of all qubits used in this experiment, including idle frequencies, average single-qubit gate errors, qubit lifetimes, and spin-echo times, are provided in Table~\ref{tab:dev_para_hamiltonian}.

\begin{table*}[!htb]
    \centering
    \addtolength{\tabcolsep}{+2pt}
\caption{\small \textbf{Parameters of processor \MakeUppercase{\romannumeral 1} and processor \MakeUppercase{\romannumeral 2}}. $\omega^{0}_m$ is the transition frequency of $Q_m$ with zero flux bias, also known as the sweet spot. $\omega^{\text{i}}_m$ is the idle frequency of $Q_m$, where single-qubit XY rotations from the Clifford group are applied. The average single-qubit gate error $e^{\mathrm{sq}}_m$ of each qubit is measured by randomized benchmarking. To switch on the interaction, all qubits are biased to the interaction frequency ($\omega^{\mathrm{I}}/2\pi \approx 4.770$~GHz for the Floquet model, and $\omega^{\mathrm{I}}/2\pi \approx 4.375$~GHz for the Hamiltonian model), where the qubit lifetime $T_{1, m}$ and spin-echo time $T_{2, m}$ of each qubit are measured. 
}

\begin{tabular}{l|c|cc|ccc}
        \hline
        \hline
        Processor \MakeUppercase{\romannumeral 1} & $\omega^{0}_m / 2\pi$ (GHz) & $\omega^{\text{i}}_m / 2\pi$ (GHz) & $e^{\mathrm{sq}}_m$ ($\%$) & $T_{1, m}$ ($\mu$s) & $T_{2, m}$ ($\mu$s) \\
        \hline
        $Q_{1 }$&   4.828&     4.790&    0.23&      84.5&     17.0\\
        $Q_{2 }$&   4.972&     4.660&    0.17&      51.9&     13.4\\
        $Q_{3 }$&   5.025&     4.748&    0.18&      119.7&     12.9\\
        $Q_{4 }$&   5.077&     4.842&    0.24&      114.9&     13.5\\
        $Q_{5 }$&   5.225&     4.800&    0.31& 
        79.8&     13.4\\
        \hline
        Average &       -&         -&    0.23&      90.2&     14.0\\
        \hline
\end{tabular}

\begin{tabular}{l|c|cc|ccc}
        \hline
        \hline
        Processor \MakeUppercase{\romannumeral 2} & $\omega^{0}_m / 2\pi$ (GHz) & $\omega^{\text{i}}_m / 2\pi$ (GHz) & $e^{\mathrm{sq}}_m$ ($\%$) & $T_{1, m}$ ($\mu$s) & $T_{2, m}$ ($\mu$s) \\
        \hline
        $Q_{1 }$&   4.578&     3.982&    0.40&      24.0&     16.5\\
        $Q_{2 }$&   4.670&     4.306&    0.42&      26.5&     19.5\\
        $Q_{3 }$&   4.639&     3.996&    0.40&      30.7&     13.3\\
        $Q_{4 }$&   4.619&     4.326&    0.35&      17.4&     10.8\\
        $Q_{5 }$&   4.622&     4.029&    0.25&      25.8&     16.6\\
        \hline
        Average &       -&         -&    0.36&      24.9&     15.3\\
        \hline
\end{tabular}

\label{tab:dev_para_hamiltonian}
\end{table*}

For both models, the system Hamiltonian can be programmed to simulate an ensemble of unitary time evolution operators $\Uh\left(t\right)$ (see Eq.~1 in the main text).
An intuitive way to measure SFF is to acquire the time evolution operators by quantum process tomography (QPT)~\cite{tomographyExp}.
However, the number of experimental runs needed to measure a time evolution operator using QPT scales exponentially with system size $L$.
Therefore, developing and implementing experimental protocols to efficiently measure SFF has recently attracted great interest~\cite{ZollerSFF2020, ZollerSFF2021, Dag:2022vqb}. 
One of the most promising protocols is proposed in Ref.~\cite{ZollerSFF2021}, which is based on the randomized measurement toolbox~\cite{Elben_2022} and makes use of the statistical correlations of local random rotations available on the state-of-the-art quantum processors.

\begin{figure}[t]
    \centering
    \includegraphics[width=\textwidth]{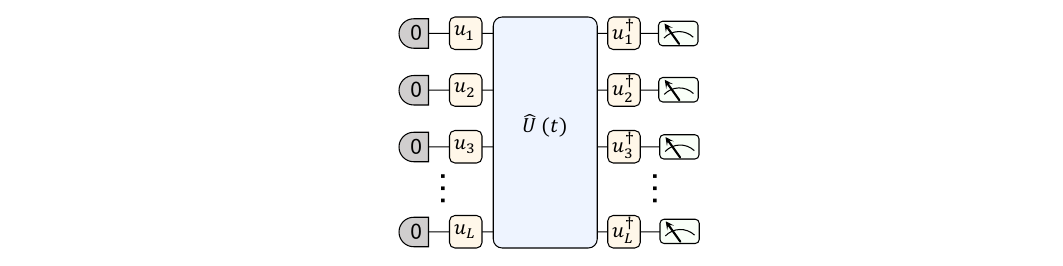}
    \caption{
        {\bf Pulse sequence for probing the SFF and pSFF dynamics with a randomized measurement protocol.}
        We start with an initial product state $|0\rangle^{\otimes L}$. Before and after the unitary time evolutions $\Uh(t)$ of the Floquet or the Hamiltonian models, we apply random Clifford gates and their inverses, respectively. Finally, we measure all qubit simultaneously for $2^{L}$ occupation probabilities.
    }
    \label{fig: SFF_rand_meas_circuit}
\end{figure}

The experimental protocol for measuring SFF using a randomized measurement toolbox is described as follows (also see Fig.~\ref{fig: SFF_rand_meas_circuit}).
We first initialize the system in the product state $|0\rangle^{\otimes L}$, and apply on-site high-fidelity random unitary gates $u_m$ that are independently drawn from the single-qubit Clifford gates.
%
%
Secondly, we tune all qubits to be in resonance at the interaction frequency $\omega^{\mathrm{I}}$ and evolve the system with $\Uh(t)$ which is generated with either a time-periodic (the Floquet model) or a time-independent (the Hamiltonian model) Hamiltonian.
%
Then, the adjoint local random unitaries $u_m^{\dagger}$ are applied on each qubit at its own idle frequency. 
Finally, we measure all qubits in the $z$-basis, and obtain a bit string of the form $\mathbf{s}=(s_1,\ldots,s_L)$ where $s_m \in \{0,1\}$ denotes the measurement result on site $m$. 
For each realization of the unitary gates and the time evolution operator, the above procedures are repeated $3000$ times and a readout-error mitigation method~\cite{PhysRevLett.118.210504} is adopted to collect the probability of each bit string $\mathbf{s}$, denoted $P_{r}\left(\mathbf{s}\right)$.
The SFF of a quantum many-body system then reads~\cite{ZollerSFF2021}
\begin{equation}
K\left(t\right)
= \frac{1}{R} 
\sum_{r=1}^R
\sum_{\mathbf{s}}
P_{r}\left(\mathbf{s}\right)\left(-2\right)^{- |\mathbf{s}|} \; ,
\end{equation}
where $\left|\mathbf{s}\right|$ is defined as $\left|\mathbf{s}\right| := \sum_{m} s_{m}$ and $R$ is the realization number.
The estimated value for pSFF $K_A(t)$ can be calculated by projecting $P_{r}\left(\mathbf{s}\right)$ onto subsystem A for $P_{r}\left(\mathbf{s}_A\right)$
\begin{equation}
K_A\left(t\right)
= \frac{1}{R} 
\sum_{r=1}^R
\sum_{\mathbf{s}_A}
P_{r}\left(\mathbf{s}_A\right)\left(-2\right)^{- |\mathbf{s}_A|} \; .
\end{equation}

\section{Error mitigation and estimation of the plateau times}

For both the Floquet and Hamiltonian models, the experimental plateau values of the SSF seem to be lower than the theoretical predictions, which is mainly due to the qubit decoherence. Consequently, in Fig.~2 and 3 of the main text, we provide numerical simulations with decoherence taken into account as more suitable references. These noisy numerics are acquired by describing the dynamics of the system with the Lindblad master equation given by
\begin{equation}
    \dot{\rho}\left(t\right) = -i \left[\hat{H}\left(t\right), \rho\left(t\right)\right]    
    + \sum_{n} \sum_{m} \left(2 \hat{C}_{n, m}^{} \rho\left(t\right) \hat{C}_{n, m}^{\dagger} - \rho\left(t\right) \hat{C}_{n, m}^{\dagger} \hat{C}_{n, m}^{} - \hat{C}_{n, m}^{\dagger} \hat{C}_{n, m}^{} \rho\left(t\right)\right),
\end{equation}
where the jump operator $ \hat{C}_{n, m}^{}$ characterize the $n$-th kind of noise acting on the $m$-th qubit.
In our specific context, we include the amplitude damping noise and dephasing noise with jump operators given by
\begin{equation}
    \begin{split}
        C_{1, m} &= \sqrt{\frac{1}{T_{1, m}}} \hat{\sigma}_m^-, \\ 
        C_{2, m} &= \sqrt{\frac{2}{T_{2, m}}} \hat{\sigma}_m^+ \hat{\sigma}_m^-, \\
    \end{split}
\end{equation}
respectively. The values of qubit lifetimes $T_{1, m}$ and spin-echo times $T_{2, m}$ are provided in Table~\ref{tab:dev_para_hamiltonian}. 
As shown in the main text, the noisy numerical results of the SFF, denoted by $K_{\mathrm{dec}}$ (and $K_{A, \mathrm{dec}}$), are in close agreement with the experimental results $K_{\mathrm{exp}}$ (and $K_{A, \mathrm{exp}}$).

\begin{figure*}[htbp]
    \centering
    \includegraphics[width=\linewidth]{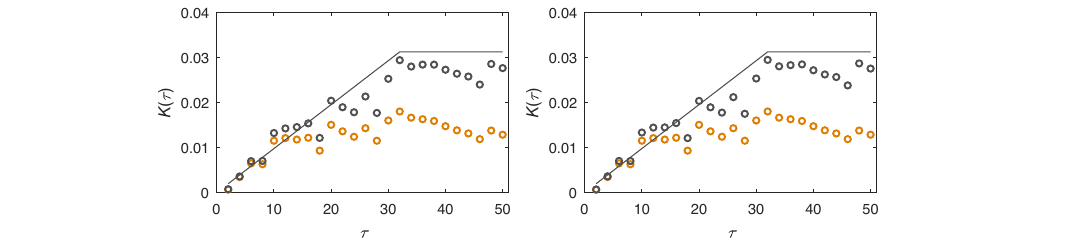}
    \caption{{\bf Error-mitigated SFF dynamics for the Floquet model.}
    Two different error mitigation schemes are implemented using the experimental SFF values (orange dots, both panels), which gives $K^{(1)}_{\mathrm{em}}(\tau)$ (gray dots, left panel) and $K^{(2)}_{\mathrm{em}}(\tau)$ (gray dots, right panel). The prediction of the SFF according to random matrix theory is shown as the solid lines.}
    \label{fig:error-mitigation}
\end{figure*}

To better visualize the ramp-plateau and dip-ramp-plateau behaviors, we propose and implement two different kinds of error mitigation schemes. The first scheme is to rescale the experimental results according to
\begin{equation}
\begin{split}
    K^{(1)}_{\mathrm{em}}\left(t\right)
    &= \frac{K_{\mathrm{sim}} \left(t\right)}{K_{\mathrm{dec}} \left(t\right)}
    K_{\mathrm{exp}} \left(t\right) \; ,
\end{split}
\end{equation}
where $K_{\mathrm{sim}}$ is the numerical results without decoherence.
The second scheme is outlined in Ref.~\cite{ZollerSFF2021} for analysis of the Floquet model, with the assumption of errors only occurring through the depolarization channel.
The resulting formula is
\begin{equation}
\begin{split}
    K^{(2)}_{\mathrm{em}}(t) &= \frac{K_{\mathrm{dec}}\left(t\right)-\left(1-\alpha\left(t\right)\right)/N^2}{\alpha(t)} \; ,
\end{split}
\end{equation}
where $\alpha\left(t\right)=\sqrt{\left(N\mathrm{Tr}\left(\rho_{\mathrm{dec}}\left(t\right)^{2}\right) - 1\right)/\left(N-1\right)}$ is determined by numerical simulations with decoherence.
As shown in Fig.~\ref{fig:error-mitigation}, two schemes give quite similar results for the Floquet model.
In the main text, we only provide the error-mitigated results using the first scheme, that is, $K^{(1)}_{\mathrm{em}}\left(t\right)$.

To extract the plateau times of the SFF for the Hamiltonian model (Fig.~3 of the main text), we numerically fit the error-mitigated data $K^{(1)}_{\mathrm{em}} \left(t\right)$ using the formula $f\left(t; \; \mu, \thei\right) = \mu K_{\mathrm{GOE}}\left(t; \; \thei\right)$, where $\mu$ and $\thei$ are two fitting parameters.
In each inset of Fig.~3(c)-(f) of the main text, we plot $K^{\prime}\left(t\right) = \mu^{-1} N K^{(1)}_{\mathrm{em}} \left(t\right)$ together with the corresponding fitting curve $N K_{\mathrm{GOE}}\left(t; \; \thei\right)$.

\section{Experimental results of the pSFF for other system sizes}

\begin{figure*}
\centering
\includegraphics[width=\linewidth]{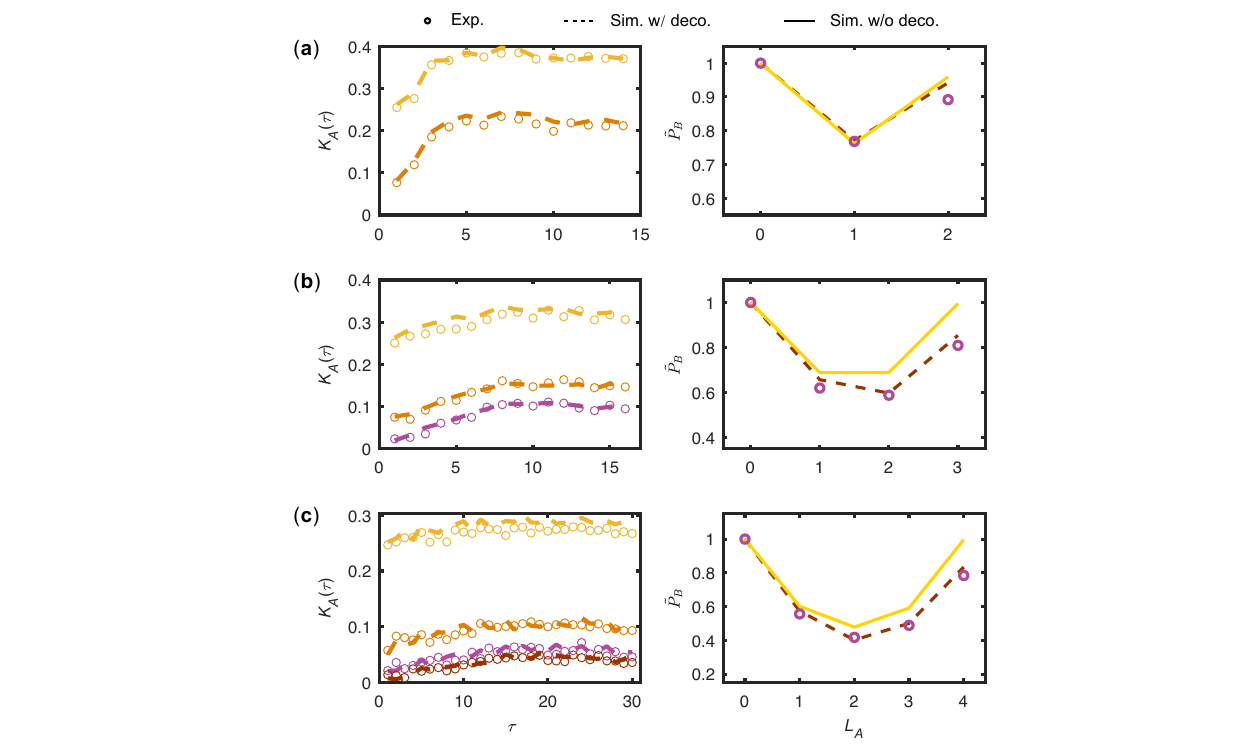}
    \caption{{\bf Numerical and experimental results of the pSFF for the Floquet model ($L=2$, $3$, and $4$).} 
    All experimental results are in close agreement with the corresponding numerical simulations with decoherence.
    By comparing the simulation results with and without decoherence, we observe that the asymmetry of the subsystem purity $\tilde{P}_{B}$ is mainly due to the decoherence.
    }
\label{Fig_PSFF1}

\end{figure*}

\begin{figure*}
\centering
\includegraphics[width=\linewidth]{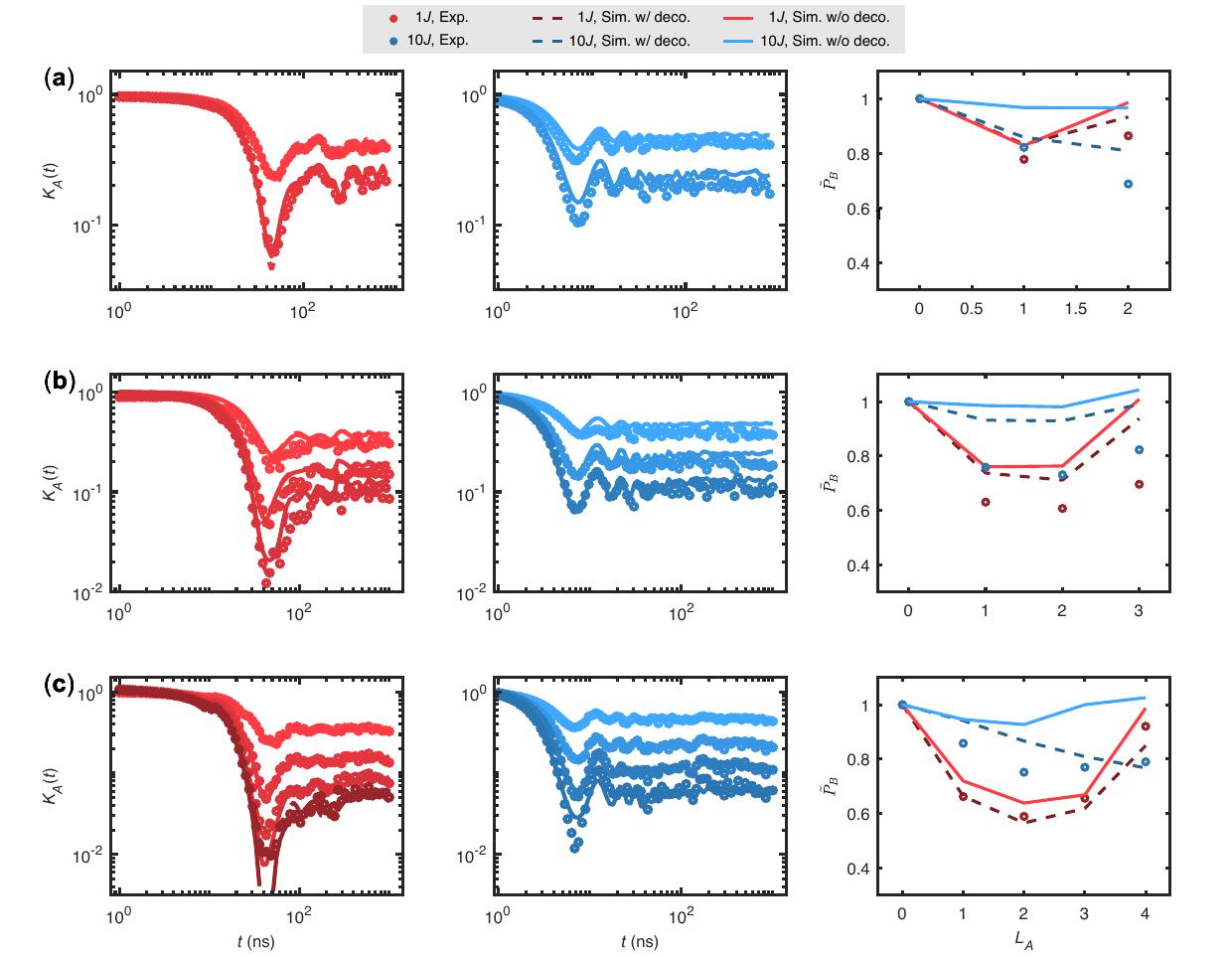}
    \caption{{\bf Numerical and experimental results of the pSFF for the Hamiltonian model ($L= 2$, $3$, and $4$).}
    By comparing the simulation results with and without decoherence, we observe that the asymmetry of the subsystem purity $\tilde{P}_{B}$ is mainly due to the decoherence.
    Importantly, the difference in the pSFF between the chaotic and prethermal many-body localized regimes is obvious for any subsystem size $L_A$.
    }
    \label{Fig_PSFF2}
\end{figure*}

Here, in Fig.~\ref{Fig_PSFF1} and \ref{Fig_PSFF2}, we provide additional experimental results of the pSFF data $K_{A}$ together with the corresponding purity $\tilde{P}_{B}$ ($L=2$, $3$, and $4$), which are not shown in the main text. 
Importantly, we observe that the plateau time of pSFF is representative and roughly independent of the subsystem size $L_A$, which is also equal to the plateau time of the whole system, both in the Floquet and the Hamiltonian models.

\end{document}